\documentstyle[twocolumn,aps]{revtex}

\def\plb{Phys. Lett. B~}
\def\pc{Phys. C~}
\def\mpla{Mod. Phys. Lett. A~}

\begin{document}
\draft
\preprint{hep-ph/9909293\\}
\title{A mass matrix for atmospheric, solar, and LSND neutrino
oscillations\\}
\author{James M. Gelb\cite{gelb}\\}
\address{
Department of Physics\\
University of Texas at Arlington\\
Arlington, Texas 76019\\
}
\author{S. P. Rosen\cite{rosen}\\}
\address{
U.S. Department of Energy\\
Germantown, Maryland 20874\\
}
\date{\today}
\maketitle

\begin{abstract}
We construct a mass matrix for the four neutrino flavors, three
active and one sterile, needed to fit oscillations in all three
neutrino experiments: atmospheric, solar, and LSND, simultaneously.
It organizes the neutrinos into two doublets whose central values
are about $1~{\rm eV}$ apart, and whose splittings are of the order of
$10^{-3}~{\rm eV}$. Atmospheric neutrino oscillations are described as
maximal mixing within the upper doublet, and solar as the same within
the lower doublet. Then LSND is a weak transition from one doublet to
the other. We comment on the Majorana versus Dirac nature of the active
neutrinos and show that our mass matrix can be derived from an $S_2
\times S_2$
permutation symmetry plus an equal splitting rule.
\end{abstract}

\narrowtext

Neutrinos produced by the interaction of cosmic rays with the Earth's
upper
atmosphere provide the strongest evidence
for neutrino oscillations\cite{atm_experiment}, with $\nu_{\mu}
\rightarrow
\nu_{\tau}$
as the favored flavor transition\cite{atm_theory}. If the additional
evidence from
solar\cite{solar_evidence} and LSND\cite{lsnd_evidence} experiments is
also
confirmed, then it will be
necessary to introduce a fourth light neutrino, a so-called
``sterile neutrino" $\nu_s$ in addition to the standard electron-,
muon-, and tau-neutrinos to account for all the data\cite{kayser_etal}.
The question then arises as to the mass spectrum and mixing
scheme for these four particles.

In a two-flavor oscillation scenario, the atmospheric data suggest
maximal mixing with mass difference $\Delta m^2 \approx 3\times
10^{-3}~{\rm eV}^2$\cite{two_flavor}. Of the three types of solution
for the
solar neutrino
data, there are two, namely the large angle MSW (LMSW) and the
``just-so"
{\it in vacuo} ones, which require close to maximal
mixing\cite{msw_justso};
while the third,
small angle MSW (SMSW) requires small mixing\cite{lsnd_smallmixing}. In
all
three cases, the
mass difference $\Delta m^2$ is much smaller than in the atmospheric
case.
By contrast, the LSND data require small mixing, but with a relatively
large
$\Delta m^2$ as compared with the atmospheric case\cite{lsnd_evidence}.

Recently, Bilenky, Giunti, Grimus, and Schwetz\cite{BGGS} have shown
that
the only way to account for these data in a four neutrino framework is
to require a mass spectrum consisting of two doublets\cite{two_doublets},
with the splitting within each doublet being
much smaller than the separation between them. Here we wish to propose
a specific realization and mass matrix in which the members of the upper
doublet are identified as maximal superpositions of $\nu_{\mu}$ and
$\nu_{\tau}$, and the members of the lower doublet are
maximal superpositions of $\nu_e$ and $\nu_s$. Atmospheric neutrino
data
can then be described as maximal oscillations between the levels
of the upper doublet, and solar neutrino data as maximal oscillations
between the levels of the lower doublet. LSND is then a weak transition
from one doublet to the other.

In adopting this point of view, we recognize that there are some
problems with the current data. The validity, or otherwise, of 
sterile neutrinos
will be tested at SNO.\cite{SNO} The main impetus for reviving the ``just-so"
solutions comes from the anomalous points at the high end of the
solar electron recoil spectrum observed at SuperKamiokande.\cite{atm_experiment}
As we have pointed out in another paper,\cite{another_look} a crucial
test for this will be the measurement of the $^7$Be neutrinos.

Our approach to the development of a mass matrix for
a two-doublet model can be illustrated with the
following two-dimensional model:
\begin{equation}
 \bar{\Psi}M_2 \Psi
  = {\left(\matrix{\bar{\psi}_a&\bar{\psi}_b\cr}\right) =
\left(\matrix{m_s &
m_k\cr
               m_k & m_s\cr}\right)}{\left(\matrix{\psi_a \cr \psi_b
\cr}\right)}~,
\end{equation}
in which the matrix $M_2$ is a linear combination of the unit $(2\times
2)$
matrix $I$ and the Pauli matrix $\sigma_x$:
\begin{equation}
 M_2 = m_sI + m_k \sigma_x~.
\end{equation}
It has eigenvalues $(m_s \pm m_k)$ and eigenstates which are
maximal mixtures of the basis states:
\begin{equation}
 \psi_{\pm} = \left(\psi_a \pm \psi_b \right)/\sqrt{2}~,
\end{equation}
and thus it will lead to maximal mixing between neutrinos $\nu_a$ and
$\nu_b$. For future reference, we note that the matrix $M_2$ is
symmetric
under the permutation group $S_2$ of the two members of the doublet,
and
that the eigenvectors $\psi_{\pm}$ are respectively even and odd
representations of $S_2$.

Now suppose we rotate $M_2$ through a small angle $(-2\delta \theta)$
about the y-axis:
\begin{eqnarray}
&\exp&(+i\sigma_y\delta\theta)M_2\exp(-i\sigma_y\delta\theta)
\nonumber\\
&=& m_sI + m_k \sigma_x\cos2\delta\theta + m_k =
\sigma_z\sin2\delta\theta
\nonumber\\
&=& {\left(\matrix{m_s + m_k\sin2\delta\theta& =
m_k\cos2\delta\theta\cr
               m_k\cos2\delta\theta & m_s -
m_k\sin2\delta\theta\cr}\right)}~.
\end{eqnarray}
It has the same eigenvalues as the original matrix, but its eigenstates
are
also rotated through the small angle $(-2\delta \theta)$:
\begin{equation}
 \psi_{\pm}(\delta\theta) =
{\left(\matrix{\psi_+\cos\delta\theta +
       \psi_-\sin\delta\theta \cr
       \psi_+\sin\delta\theta - \psi_-\cos\delta\theta\cr}\right)}~,
\end{equation}
and so it leads to small mixing oscillations between $\psi_+$ and
$\psi_-$.

Guided by this analysis, we propose a four-flavor mass matrix which
we construct by replacing $m_s$ and $m_k$
in the rotated form of $M_2$ by $(2\times 2)$ matrices:
\begin{eqnarray}
m_s &\rightarrow& M~,~~
M  = {\left(\matrix{m_s & m_d \cr
                   m_d & m_s \cr}\right)}~; \\
m_k &\rightarrow& K~,~~
K  = {\left(\matrix{ m_k & 0\cr
                    0   & m_k\cr}\right)}~.
\end{eqnarray}
Our model then takes the form
\begin{eqnarray}
&\bar{\Psi}&M_4 \Psi = \\
 &&\left(\matrix{\bar{\Psi}_a & \bar{\Psi}_b\cr}\right)
{\left(\matrix{M + K\sin2\delta\theta& K\cos2\delta\theta\cr
               K\cos2\delta\theta & M - K\sin2\delta\theta\cr}\right)}
                {\left(\matrix{\Psi_a \cr \Psi_b
\cr}\right)}~,\nonumber
\end{eqnarray}
where $\Psi_a$ and $\Psi_b$ are now two-dimensional column vectors:
\begin{equation}
  {\left(\matrix{\Psi_a & \Psi_b}\right)} =
{\left(\matrix{\psi_{a1} & \psi_{b1}\cr
                 \psi_{a2} & \psi_{b2} \cr}\right)}~.
\end{equation}

Next we rotate $M_4$ and $\Psi$ into the forms:
\begin{eqnarray}
 M_4 & \rightarrow & \tilde{M}_4 = {\left(\matrix{M& K\cr
                    K & M\cr}\right)}~, \\
 \Psi &\rightarrow & \Phi = {\left(\matrix{\Phi_a \cr \Phi_b
\cr}\right)} =
{\left(\matrix{\cos\delta\theta& -\sin\delta\theta\cr
               \sin\delta \theta & ~\cos\delta\theta\cr}\right)}
                         {\left(\matrix{\Psi_a \cr \Psi_b
\cr}\right)}~.
\end{eqnarray}
For future reference, we note that $\tilde{M}_4$ is symmetric under the
permutation group $\tilde{S_2}$ which interchanges the two doublet
pairs.

Now we rotate $\tilde{M}_4$ and $\Phi$ into:
\begin{eqnarray}
 \tilde{M}_4 & \rightarrow & {\left(\matrix{M+K & 0\cr
                    0 & M - K\cr}\right)}~, \\
 \Phi & \rightarrow & {1 \over \sqrt 2}{\left(\matrix{~\Phi_a & +
&\Phi_b
\cr
-\Phi_a & + & \Phi_b
\cr}\right)}~.
\end{eqnarray}

We now have to diagonalize the $(2\times 2)$ matrices $(M\pm K)$, where
\begin{equation}
 M \pm K = {\left(\matrix{m_s \pm m_k & m_d \cr
                    m_d & m_s \pm m_k\cr}\right)}~,
\end{equation}
which have eigenstates ${(\Phi_a + \Phi_b)/\sqrt{2}}$ and
${(-\Phi_a + \Phi_b)/\sqrt{2}}$ respectively.
The eigenvalues of
$(M + K)$ are:
\begin{equation}
 M^+_{\pm} = m_s + m_k \pm m_d~,
\end{equation}
and those of $(M - K)$ are:
\begin{equation}
 M^-_{\pm} = m_s - m_k \pm m_d~.
\end{equation}
Thus we have two doublets whose mean masses are separated by $2m_k$,
and whose splittings are both given by $2m_d$. The upper and lower
components of $(\Phi_a + \Phi_b)/\sqrt{2}$ are maximally
mixed, as are those of $(-\Phi_a + \Phi_b)/\sqrt{2}$. Finally,
the eigenstates of $(M + K)$ are weakly mixed with those of $(M - K)$
via the relation between $\Phi$ and $\Psi$ in Eq.~11 above.

We identify $(M + K)$ and its eigenstates with the atmospheric
neutrino oscillations between $\nu_{\mu}$ and $\nu_{\tau}$,
and so the squared mass difference may be written
\begin{eqnarray}
\Delta_A & = & (m_s + m_k + m_d)^2 - (m_s + m_k - m_d)^2 \nonumber\\
         & = & 4(m_s + m_k)m_d~.
\end{eqnarray}
Similarly, we identify $(M - K)$ and its eigenstates with
solar neutrino oscillations between $\nu_e$ and $\nu_s$, and so
\begin{eqnarray}
\Delta_S & = & (m_s - m_k + m_d)^2 - (m_s - m_k - m_d)^2 \nonumber\\
         & = & 4(m_s - m_k)m_d~.
\end{eqnarray}
For reasons which will become apparent below, we write
\begin{eqnarray}
m_s  & = & m_0 + \epsilon~, \\
m_k & = & m_0 - \epsilon
\end{eqnarray}
and so
\begin{equation}
{\epsilon \over m_0} = {\Delta_S\over \Delta_A}~.
\end{equation}
Since $\Delta_A$ is much greater than $\Delta_S$, as discussed below,
we conclude that $\epsilon$ is much smaller than $m_o$, and that
$m_s$ is only marginally greater than $m_k$:
\begin{equation}
{m_s \over m_k} \approx (1 + 2\epsilon) \approx 1 + 2{\Delta_S\over
\Delta_A}~.
\end{equation}

For LSND, we assume that the $\bar{\nu}_{\mu} \rightarrow \bar{\nu}_e$
oscillation is dominated by the transition from the lower eigenvalue
of $(M + K)$ to the upper eigenvalue of $(M - K)$:
\begin{eqnarray}
\Delta_L & = & (m_s + m_k - m_d)^2 - (m_s - m_k + m_d)^2 \nonumber\\
& = &
4(m_k - m_d)m_s~, \end{eqnarray}
and so
\begin{eqnarray}
8m_k m_s  &=&  8(m_0^2 - \epsilon^2) \nonumber\\
          &=&  2\Delta_L + \Delta_A + \Delta_S~.
\end{eqnarray}
Since $\Delta_L$ is much bigger than either $\Delta_A$ or $\Delta_S$,
it follows that:
\begin{equation}
2m_0 \approx \sqrt{\Delta_L}~ \left(1 + {\Delta_A \over
4\Delta_L}\right)~.
\end{equation}
We then find that $m_d$ is much smaller than $m_0$:
\begin{equation}
2m_d \approx {\Delta_A \over 2\sqrt{\Delta_L}}~{ \left( {1 - {\Delta_A
\over
4\Delta_L}} \right) }~.
\end{equation}

To gain a sense of the magnitude of the mass matrix elements,
we assume the following values for the observed mass-squared
differences:
\begin{eqnarray}
\Delta_L & \approx& 1~{\rm eV}^2~, \nonumber \\
\Delta_A & \approx& 3\times 10^{-3}~{\rm eV}^2~, \nonumber \\
\Delta_S & \approx& 10^{-5}~{\rm eV}^2~,
\end{eqnarray}
and so the ratios of mass-squared differences are all the same, namely
\begin{equation}
{\Delta_A \over \Delta_L} \approx {\Delta_S \over \Delta_A} \approx
3\times
10^{-3}~.
\end{equation}
It is interesting to note that, for the above value of $\Delta_L$,
this is also the value of the weak mixing angle between upper and lower
doublets needed to fit the LSND data\cite{lsnd_evidence}:
\begin{equation}
\sin^2 2\delta \theta \approx 3\times 10^{-3}~.
\end{equation}

The large parameter in the mass matrix, $m_0$, is close to $0.5~{\rm =
eV}$:
\begin{equation}
2m_0 \approx 1.001~{\rm eV}~;
\end{equation}
and the small parameters, $\epsilon$ and 2$m_d$, are much smaller
and roughly equal to one another,
\begin{equation}
\epsilon \approx 1.5 \times 10^{-3}~{\rm eV}~,~~
2m_d \approx 1.5 \times 10^{-3}~{\rm eV}~.
\end{equation}
Thus the upper doublet, corresponding to $\nu_{\tau}$ and $\nu_{\mu}$,
has a central value of $1.001~{\rm eV}$ and a
splitting of $1.5 \times 10^{-3}~{\rm eV}$, while the lower doublet,
corresponding to
$\nu_e$ and $\nu_s$,
has an almost zero central value, $3\times 10^{-3}~{\rm eV}$, with
the same splitting as the upper one.

We have not considered the Majorana versus Dirac nature of the
four neutrinos and the constraints from no-neutrino double beta
decay\cite{double_beta}. If the three active ones are all Majorana
particles, then
the sum of their masses times CP phase must not exceed the current
bound of $0.2-0.6~{\rm eV}$\cite{cp_bound}. In the above example, this
is
most easily
achieved by giving the members of the upper doublet opposite CP
phases, which make them ``Pseudo-Dirac" neutrinos because of the
small mass difference $2m_d$. Whatever phase is assigned to the
active member of the lower doublet, the sum of masses times phase
will not exceed $6 \times 10^{-3}$ eV, well within the experimental
limit\cite{experimental_limit}.

We may now ask whether the mass matrix $M_4$ can be derived from a
symmetry principle. As we have noted above, the case of maximal mixing
among the two members of a doublet corresponds to the permutation
symmetry $S_2$
between them. Likewise the general structure of $M_4$ involves the
permutation
symmetry $\tilde{S_2}$ between the two doublets. It is not difficult to
show
that the most general $4 \times 4$ matrix $\it{H_4}$ which is invariant
under
$S_2 \times \tilde{S_2}$ is given by:
\begin{eqnarray}
\it{H_4} = {\left(\matrix{\it{X} & \it{Y}\cr
                    \it{Y} & \it{X}\cr}\right)}~.
\end{eqnarray}
The $(2 \times 2)$ submatrices $\it{X}, \it{Y}$ are both of the same
$S_2$
symmetric form as $M_2$ above.

Comparing $M_4$ with $H_4$, we see that it is of exactly the same form
except
that the off-diagonal submatrix $K$ is a multiple of the unit $(2 \times
2)$ matrix
whereas $\it{Y}$ can have an off-diagonal matrix element. Physically,
the absence
of an off-diagonal matrix element in $K$ means that the splitting
between the members
of the upper doublet is exactly the same as that between the members of
the lower
doublet---an ``equal splitting" rule.

In conclusion, we have constructed a mass matrix which can
simultaneously accommodate all three indications for neutrino
oscillations. Its particular structure as a direct product
of $(2\times 2)$ matrices can be derived from an underlying $S_2 \times
\tilde{S_2}$
symmetry plus an equal splitting rule. It may be interesting to
speculate that
this symmetry might in turn be a subgroup of a
larger permutation symmetry, for example $S_4$, and that the larger
symmetry can be
used to distinguish between the active and sterile neutrinos. For
example, the three
active neutrinos could
belong to a triplet with respect to an $S_3$ subgroup of the larger
group,
while the sterile neutrino is a singlet.

We recognize that large mixing between a sterile neutrino and 
the electron-neutrino in the Solar Neutrino Problem can disturb 
Big Bang Nucleosynthesis,\cite{BBN} and we have no ready solution 
for this problem. Whether Big Bang Nucleosynthesis 
can accommodate 3 or 4 light
neutrino degrees of freedom will depend crucially on the amount
of primordial deuterium in the universe;
at the moment this is not well determined.\cite{burles_olive}
We do, however, regard the existence, or 
non-existence, of a sterile neutrino to be an experimental question 
which will eventually be settled by the observation of the 
neutral-current interactions of solar neutrinos, as in the SNO
experiment.\cite{SNO}

We are indebted to Hamish Robertson for asking a question
which sparked this investigation.


\begin{references}

\bibitem[*]{gelb} Email address:~~gelb@alum.mit.edu.

\bibitem[{\dagger}]{rosen} Email address:~~Peter.Rosen@oer.doe.gov.

\bibitem{atm_experiment}  SuperKamiokande Collaboration, Y. Suzuki, in
{\it Neutrino 98},
Proceedings of the XVIII International Conference on Neutrino Physics
and Astrophysics, Takayama, Japan, 1998, edited by Y. Suzuki
and Y. Totsuka [Nucl. Phys. B (Proc. Suppl.) (in press)];
Y. Fukuda {\it et al.}, \prl {\bf 81}, 1158 (1998); {\it ibid}, 
{\bf 81}, 4279(E)
(1998);
T. Mann in {\it Lepton-Photon 99}, SLAC, August (1999).
For an extensive discussion of neutrino mass matrices, 
see R. N. Mohapatra (1999), hep-ph/9910365.

\bibitem{atm_theory} For a review,
see P. Fisher, B. Kayser, and K. S. McFarland,
{\it Annual Reviews of Nuclear and Particle Science} {\bf 49} (1999),
hep-ph/9906244;
V. Gribov and B. Pontecorvo, Phys. Lett. {\bf 28B},
493 (1969);
V. Barger, K. Whisnant, and R. Phillips,
\prd {\bf 24}, 538 (1981);
S. L. Glashow and L. M. Krauss,
\plb {\bf 190}, 199 (1987);
V. Barger, R. Phillips, and K. Whisnant, \prd {\bf 43}, 1110
(1991); {\it ibid}, \prl {\bf 69}, 3135 (1992); P. I. Krastev and S.
Petcov,
\prd {\bf 53}, 1665 (1996); J. N. Bahcall, P. I. Krastev, and E. Lisi,
\prc {\bf 55}, 494 (1997);
G. L. Fogli, E. Lisi, and D. Montanino, \prd {\bf 56}, 4374 (1997);
A. J. Baltz, A. S. Goldhaber, and M. Goldhaber, \prl
{\bf 81}, 5730 (1998).

\bibitem{solar_evidence} S. Suzuki in {\it Lepton-Photon 99}, SLAC,
August
(1999);
SAGE Collaboration, V. Gavrin {\it et al.}, in {\it Neutrino
96}, Proceedings of the XVII International Conference on Neutrino
Physics and Astrophysics, Helsinki, Finland, 1996,
edited by K. Huitu, K. Enqvist, and
J. Maalampi (World Scientific, Singapore, 1997), p.14; J. N.
Abdurashitov {\it et al.}, \prl {\bf 77}, 4708 (1996);
GALLEX Collaboration,
P. Anselmann {\it et al.}, \plb {\bf 342}, 440
(1995); W. Hampel {\it et al.}, {\it ibid}, {\bf 388}, 364 (1996); T.
Kirsten, in {\it Neutrino 98}\cite{atm_experiment}.

\bibitem{lsnd_evidence} L. DiLella in {\it Lepton-Photon 99}, SLAC,
August
(1999);
C. Athanassopoulos {\it et al.}, \prc {\bf 58}, 2511 (1998).

\bibitem{kayser_etal} P. Fisher, B. Kayser, and K. S. McFarland,
{\it Annual Reviews of Nuclear and Particle Science} {\bf 49} (1999),
hep-ph/9906244.
For a phenomenological analysis of existing solar neutrino data, 
see J. N. Bahcall, P. Krastev, and A. Y. Smirnov, 
\prd {\bf 58}, 096016 (1998).

\bibitem{two_flavor} For example, V. Gribov and B.
Pontecorvo\cite{atm_theory}.

\bibitem{msw_justso} S. L. Glashow and L. M. Krauss\cite{atm_theory};
S. L. Glashow, Peter J. Kernan, and L. M. Krauss, \plb {\bf 445}, 412
(1999).

\bibitem{lsnd_smallmixing} J. M. Gelb and
S. P. Rosen, \prd {\bf 60}, 011301 (1999); P. I. Krastev and
S. T. Petcov, Nucl. Phys. B {\bf 449},
605 (1995);
S. P. Mikheyev
and A. Yu. Smirnov, \plb {\bf 429}, 343 (1998);
J. N. Bahcall, S. Basu, and
M. H. Pinsoneault, \plb {\bf 433}, 1 (1998).

\bibitem{BGGS} S. M. Bilenky, C. Giunti, W. Grimus, and T. Schwetz,
\prd
{\bf 60}, 073007 (1999).

\bibitem{two_doublets} J. T. Peltoniemi, D. Tommasini, and J. W. F. Valle,
\plb
{\bf 298}, 383 (1993); J. T. Peltoniemi and J. W. F. Valle, Nucl. Phys.
{\bf B} 406, 409
(1993); D. O. Caldwell and R. N. Mohapatra, \prd {\bf 48}, 3259 (1993);
E. Ma and
P. Roy, {\it ibid}, {\bf 52} R4780 (1995); E. J. Chun {\it et al.}, \plb 
{\bf 357}, 608 (1995); J. J. Gomez-Cadenas and M. C. Gonzalez-Garcia,
Z. \pc {\bf 71}, 443 (1996); E. Ma, \mpla {\bf 11},
1893 (1996);
S. Goswami, \prd {\bf 55}, 2931 (1997).

\bibitem{SNO} See, for example, The Sudbury Neutrino Observatory 
Collaboration (1999), nucl-ex/9910016.

\bibitem{another_look} J. M. Gelb and S. P. Rosen, \prd {\bf 60}, 011301 (1999).

\bibitem{double_beta} R. G. H. Robertson in {\it Lepton-Photon 99},
SLAC,
August (1999); S. M. Bilenky, C. Giunti, W. Grimus, B. Kayser,
and S. T. Petcov (1999), hep-ph/9907234.

\bibitem{cp_bound} For example, R. G. H. Robertson in {\it
Lepton-Photon
99}, SLAC,
August (1999).

\bibitem{experimental_limit} L. Baudis {\it et al.} (1999),
hep-ex/9902014,
for a complete set
of references.

\bibitem{BBN} R. Barbieri and A. Dolgov, \plb {\bf 237}, 440 (1990); 
K. Enqvist, K. Kainulainen, and J. Maalampi, {\it ibid} {\bf 249}, 531 (1990); 
X. Shi and G. Fuller (1999), astro-ph/9904041.

\bibitem{burles_olive} S. Burles, K. M. Nollett, J. N. Truran, and M. S. Turner, 
\prl {\bf 82}, 4176 (1999); K. A. Olive, astro-ph/9903309 (1999).

\end{references}
\end{document}